\documentclass{aa}
\usepackage{graphics}
\usepackage{rotating}
\begin{document}
\newcommand{\mcol}[3]{\multicolumn{#1}{#2}{#3} }
\newcommand{\struut}{\rule[-2ex]{0ex}{5.2ex}}
\newcommand{\struutup}{\rule{0ex}{3.2ex}}
\newcommand{\struutdown}{\rule[-2ex]{0ex}{2ex}}
\setcounter{equation}{20}

   \thesaurus{06  
              (08.22.2 ;08.02.6 ; 08.09.2 HD 220392, HD 220391; 08.06.3 )} 
%
   \title{$\delta$ Scuti stars in stellar systems: on the variability of HD~220392 and HD~220391\thanks{
 Based on observations done at La Silla (ESO, Chile) and on data
 obtained by the Hipparcos astrometry satellite }}

   \author{P. Lampens \inst{1}, M. Van Camp \inst{1}
          \and 
           D. Sinachopoulos \inst{1,2}
          }

   \offprints{P. Lampens}

   \institute{
             Koninklijke Sterrenwacht van Belgi\"e,
             Ringlaan 3, B-1180 Brussel, Belgium\\
             email: patricia.lampens@oma.be
         \and
             National Observatory of Athens, 
	     Lophos Nymphon, GR-11810, Greece\\
            }

   \authorrunning{Lampens, P. et al.}
   \titlerunning{On the variability of HD~220392 and HD~220391}
   \date{Received date/ Accepted date}

   \maketitle

   \begin{abstract}
HD~220392 (HR~8895), the brightest member of the visual double star CCDM~23239-5349, 
is a new short-period variable bright star, probably of the $\delta$ Scuti type. 
The period analysis performed on the complete set of definitive Geneva photometry as 
well as on the data obtained at the ESO 0.5m telescope shows two periodicities 
of about 4.7 and 5.5 cycles per day (cpd) with amplitudes of 0.014 and 0.011 mag
respectively. A similar period search on the (smaller) dataset obtained for the 1 mag 
fainter B-component, HD~220391, however shows no periodicity with an amplitude 
significantly above the noise level of the data (about 0.006~mag). This difference 
in variability behaviour is discussed from the consideration that {\it both} stars 
form a common origin pair and are located in the $\delta$ Scuti instability strip.


   \end{abstract}

\section{Introduction}

The very wide double star CCDM~23239-5349 is an interesting study case of a 
pulsating star within a common origin pair or wide binary. The detailed 
investigation of the difference in variability and physical parameters between 
two components of a physical couple is particularly worthwhile when both 
stars are located in the same area of the colour-magnitude diagram, in this 
case {\it both} components are in the $\delta$ Scuti instability strip. The aim of 
such a study is to search for clues to understand what factors determine the pulsation 
characteristics such as modes and amplitudes among $\delta$ Scuti stars in general.\\ 
The Hipparcos satellite measurements confirm what was already  hinted by 
the ground-based astrometric data in the Washington Double Star Catalogue (WDS 1996.0, 
Worley \& Douglass \cite{wor97}), namely that the wide angular 
separation of 26.5 \arcsec of the system is accompanied by a very small 
relative proper motion ($\Delta\mu_{\alpha^{\ast }_{B-A}}\simeq -2.44$~milli-arcsec/yr (mas/yr), 
$\Delta \mu_{\delta_{B-A}}\simeq +1.57$~mas/yr with errors of the same order). 
The new parallaxes are furthermore compatible to better than $1.5\,\sigma$. 
This may indicate a common origin if not a true physical association (Sect.~\ref{sec:ass}).

Regular short-period light variations on a time scale of $\simeq$ 5 hr have been detected 
for the brightest component of this visual double star (Lampens \cite{lam92}).  
We describe the available observations and the reduction methods in Sect.~\ref{sec:obs}. 
The results of the period analyses are presented in Sect.~\ref{sec:res}. Also included 
is the analysis of a selection of the Hipparcos Epoch Photometry data. We discuss 
the nature of the association and of the variability in Sect.~\ref{sec:nat}. Finally we 
draw our conclusions in Sect.~\ref{sec:con} and we explain why additional observations 
for both stars of the system would be highly desirable.

\section{Observations and reductions}\label{sec:obs}

The photometric data have been gathered during three campaigns at La Silla,
Chile. For the A-component HD~220392, 7 nights of measurements were made in
June 1990, 11 nights were obtained in September 1991 and 3 in October 1992.
For the B-component HD~220391, only observations made in September 1991 and
October 1992 are available. The June 1990 and September 1991 campaigns have
been performed by P.~Lampens with the Swiss 0.7m telescope of the Geneva Observatory
while the October 1992 data were obtained with the ESO 0.5m telescope by D.~
Sinachopoulos. We have collected a total of 396 data for the brightest component HD~220392
and 245 for HD~220391. The characteristics of these data are mentioned in Table~\ref{photo}. 
Standard and additional programme stars have also been observed during these nights. 
All Geneva measurements are absolute measurements in the filters 
UBVB$_{1}$B$_{2}$V$_{1}$G obtained through a centralized reduction scheme at the 
Geneva observatory (Rufener \cite{ruf88}). 
This centralized Swiss processing has not been applied to the ESO data taken in the UBV 
photometric system. The reduction of the October data implied using a check-star 
HD~220729 [F4V, V=5.52, B-V=+0.40] whose measurements were interpolated between the two other ones. We have verified 
the constancy of this star in the Hipparcos catalogue ($H_{p}=5.6197\,$mag,$\sigma _{H_{p}}=0.0005\,$mag)
and we have fitted a 5th degree polynomial to the check-star data for each night separately. Then we have subtracted 
this polynomial from the data of both programme stars in order to suppress as well as 
possible common variations. The ESO data are thus being interpreted as differential measurements 
relative to HD 220729 only. We kept the differential data acquired at the end of the nights at relatively large airmasses 
($F_{z} >$ 1.6) though they are affected by larger noise, after some trials with various combinations. 
Our results will thus be based on the largest available datasets.
In addition we made use of the data provided in the Hipparcos Epoch Photometry Catalogue 
(ESA~\cite{esa97}).

\begin{table}[hbp]
\setlength{\tabcolsep}{1.0mm}
\caption{Photometric data for HD~220392 and HD~220391}
\label{photo}
\begin{center}
{\small
\begin{tabular}{|c|cccc|}
\hline
Identifier & Instrument & Epoch & Number & Time base \struutup\\ 
&  &[mo/yr]  & of data & [days] \struutdown\\ 
\hline
HD 220392 & Geneva & 06/90-09/91 & 124 & 464 \struutup \\ 
 & ESO & 10/92 & 272 & 9.2 \\ 
 & Gen.+ESO & 06/90-10/92 & 396 &  866 \\ 
 & Hipparcos & 11/89-03/93 & 176 &  {\it 41 months} \struutdown \\ 
\hline
HD 220391 & Geneva & 09/91 & 98 & 19.1 \struutup \\
 & ESO & 10/92 & 146 & 9.2 \\ 
 & Gen.+ESO & 09/91-10/92 & 245 &  416 \\ 
 & Hipparcos & 11/89-03/93 & 172 &  {\it 41 months} \struutdown \\ 
\hline 
\end{tabular} }
\end{center}
\end{table}

\section{Period analyses}\label{sec:res}

\subsection{HD~220392}

\subsubsection{Geneva data}

The block of 124 data for HD~220392 covers an interval of 464 days (Table~\ref{photo}).
We used the frequency step of $5.8\,10^{-5}$ cpd ($\simeq 1/20{\rm T}$) with the PERIOD98
software (Sperl \cite{spe98}). After Fourier analysis of the visual magnitudes, m$_{V}$, 
the frequencies, amplitudes and phases were improved by a least squares fit that gave a main
frequency around 4.679 cpd, the same one as previously reported by Lampens
(\cite{lam92}). The standard deviation dropped by more than 28\thinspace \% after
prewhitening for this frequency. Since the theoretically expected noise level 
of 0.006 mag for a bright constant star observed in the Geneva Photometric 
System (Rufener \cite{ruf88}) was not yet reached, a search for a second frequency 
in the prewhitened data was performed, revealing either 5.520 or 6.520 cpd.
The (1 day)$^{-1}$ ambiguity due to the spectral window in the search
for the second frequency is obvious (called "leakage effect" in Bloomfield \cite{blo76}, 
see Sect.~\ref{ssect312} below).
The second highest amplitude was found for a two-frequency fit with 5.520 cpd: 
results of the simultaneous fits are presented in Table~\ref{freq92}(b).\\
After prewhitening for the frequencies 4.679 and 5.520 cpd, the residual standard
deviation falls to 0.0085 mag, still larger than expected. However, there is very
clear evidence from the plots of the phase diagrams that the 7 data points on 
JD 2448518 have a level that is about 0.01 mag off compared to the rest of the data. 
This accounts for an extra 0.001 mag residual dispersion. A last Fourier analysis 
was done, giving 4.32 cpd and a standard deviation of 0.0073 mag after a third 
prewhitening. Evidence for this frequency is small (Sect.~\ref{ssect312}). 
Similar results are found for the Geneva m$_{U}$ and m$_{B}$ magnitudes. 
The fitted amplitudes for a two-frequency fit (preference was given to 5.520 cpd) 
are also listed in Table~\ref{freq92}(a).

\begin{table}[h]
\setlength{\tabcolsep}{0.5mm}
\caption{Results of a two-frequency fit (a) for the Geneva U and B data 
(b) for the Geneva V data and (c) for all 396 data of HD~220392 (program PERIOD98) }
\label{freq92}
\begin{center}
{\small
\begin{tabular}{|c|c|cccc|}
\hline
Set & Filter & Frequency & Semi-ampl. & Residual $\sigma $ & Reduction \struutup \\ 
 & & [cpd] & [mag] & [mag] & \% \struutdown \\ 
\hline
 (a) & U & \textit{{f$_{1}$}, 4.679} & 0.0125 & 0.01302 & 24.2 \struutup \\ 
 && \textit{{f$_{2}$}, 5.520} & 0.0125 & 0.01008 & 17.1 \struutdown \\ \hline
    & B & \textit{{f$_{1}$}, 4.679} & 0.0158 & 0.01397 & 28.0 \struutup \\ 
 && \textit{{f$_{2}$}, 5.520} & 0.0139 & 0.01048 & 18.0 \struutdown \\ \hline
 (b) & V & \textit{{f$_{1}$}, 4.679} & 0.0137 & 0.01070 & 28.7 \struutup \\ 
 && \textit{({f$_{2}$}, 6.520)} & 0.0096 & 0.00845 & 15.1 \struutdown \\ \cline{3-6}
 && \textit{{f$_{1}$}, 4.679} & 0.0128 & 0.01070 & 28.7 \struutup \\ 
 && \textit{{f$_{2}$}, 5.520} & 0.0099 & 0.00845 & 15.1 \struutdown \\ \hline
 (c) & V & \textit{{f$_{1}$}, 4.67437} & 0.0155 & 0.00911 & 40.8 \struutup \\ 
 && \textit{{f$_{2}$}, 6.52260} & 0.0097 & 0.00644 & 17.3 \struutdown \\ \cline{3-6}
 && \textit{{f$_{1}$}, 4.67439} & 0.0139 & 0.00915 & 40.6 \struutup \\ 
 && \textit{{f$_{2}$}, 5.52234} & 0.0110 & 0.00614 & 19.5 \struutdown \\ 
\hline
\end{tabular} }
\end{center}
\end{table}

\subsubsection{\protect\bigskip ESO and Geneva data}\label{ssect312}

\begin{figure*}[ht] 
 \resizebox{\hsize}{!}{\includegraphics[height=9cm,angle=0]{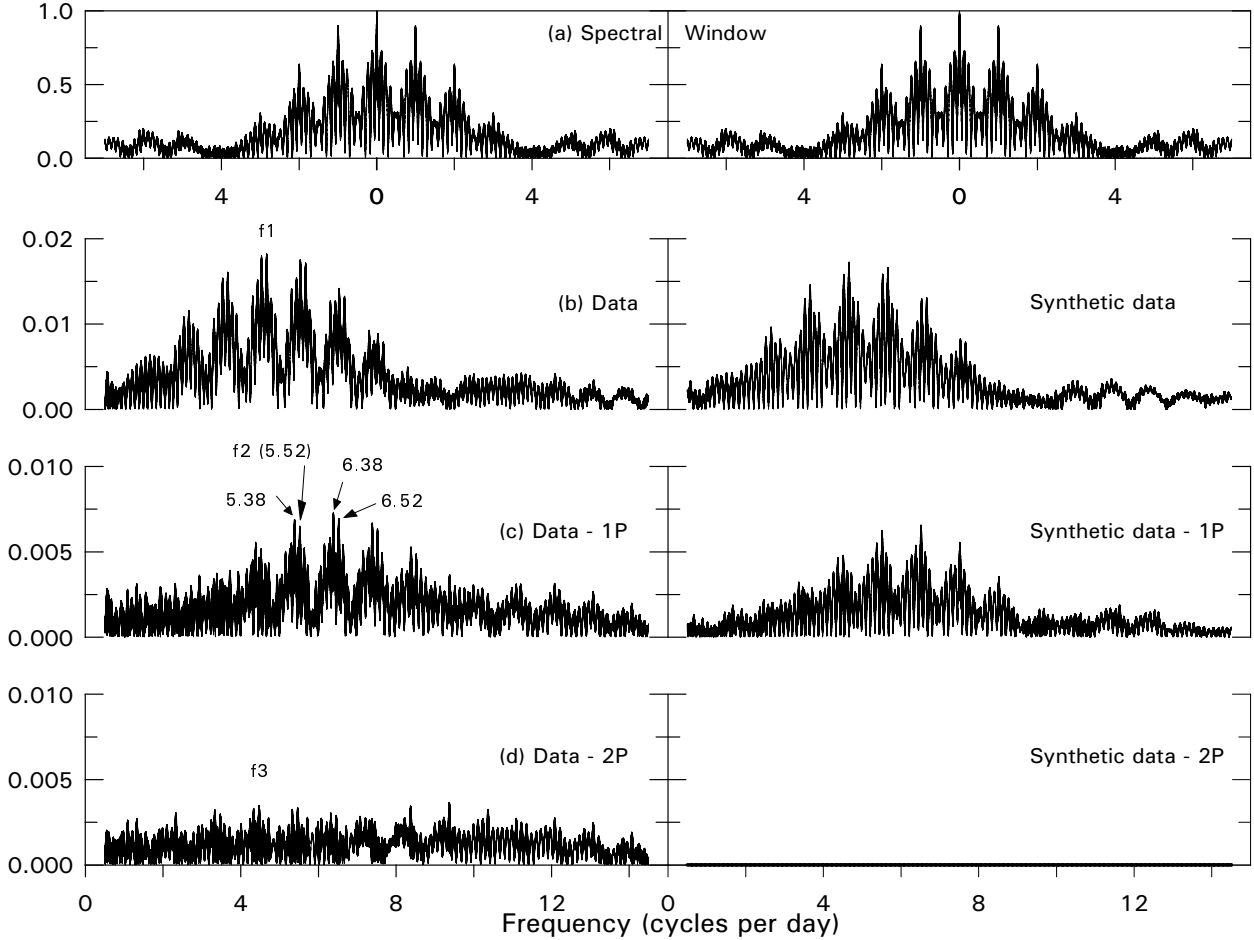}}
 \caption{Amplitude spectra for the September 91/October 92 data. Notice the strong leakage 
 effect that produces the 5.38, 6.38 and 6.52 cpd frequencies on both real
 and synthetic data}
 \label{specalia}
\end{figure*} 

The combination of data was done in the V~filter only, as the signal-to-noise
ratio of the ESO B~data is not as good as that of the V~data and because 
there are fewer ESO U~data.
To this effect we adjusted for both stars the mean V values of the ESO (differential) 
data to the corresponding mean Geneva V magnitudes of the September 91 set.
Thus adding the ESO data taken in October 1992 to the Geneva observations, a
total of 396 V data with a time base of 866 days is available. We have tried
different combinations with the datasets that confirm the results obtained
with the Geneva data (Table~\ref{freq92b}). The number of nights (Nights) and
the resolution per dataset (Resol.) are given as well. After prewhitening for 4.67 cpd, a new
spectral analysis gives peaks at 6.52 or 6.38 cpd for all datasets, 
except for the complete set of 21 nights which gives 5.52 cpd. These
frequencies are shown in brackets on Table~\ref{freq92b}. Among them, we have
preferred 5.52 cpd for three reasons. First, it is the second dominant
frequency in the largest dataset. Second, amplitudes for $f_{2}$ given by the 
least squares fit are always larger with 5.52 than with 6.52 cpd while standard 
deviations of the residuals are generally smaller after prewhitening 
with 5.52 cpd than in the case with 6.52 cpd. Third, an analysis made with the 
synthetic wave $0.0136\sin(2\pi t\,4.664)+0.0092\sin (2\pi t\,5.52)$ using 
the time window of September 1991/October 1992 gives as main frequencies 4.664 
and 6.52 cpd, occulting the one of 5.52 cpd. This phenomenon is illustrated by 
Fig.~1 and is due to the "leakage effect" induced by the night/day alternation. 
Using these same arguments, we found that the 
frequency of 6.38 cpd as observed with the October and September/October datasets 
is also due to leakage, caused by a gap in the October 1992 campaign.

\begin{figure}[ht]
\label{pha11}
 \resizebox{\hsize}{!}{\includegraphics[height=6.5cm,angle=270]{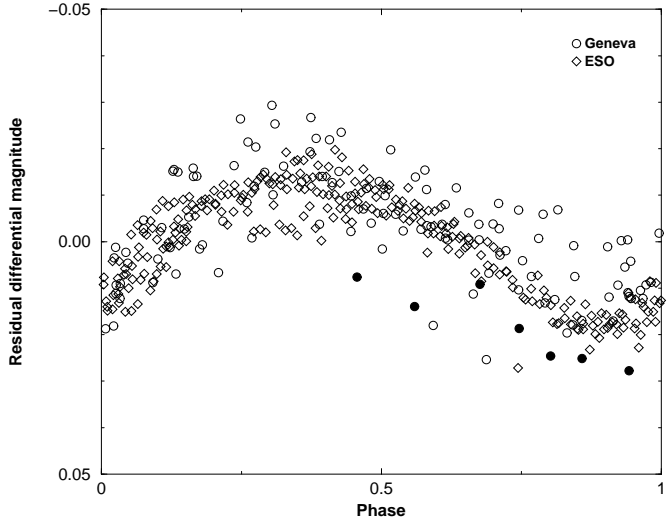}}
 \caption{Phase diagram for the HD~220392-data against the frequency of 4.67439 cpd 
 (after removal for the 5.52 cpd variation). Filled symbols represent
 the data on JD 2448518 }
\end{figure}

\begin{figure}[ht]
\label{pha22}
 \resizebox{\hsize}{!}{\includegraphics[height=6.5cm,angle=270]{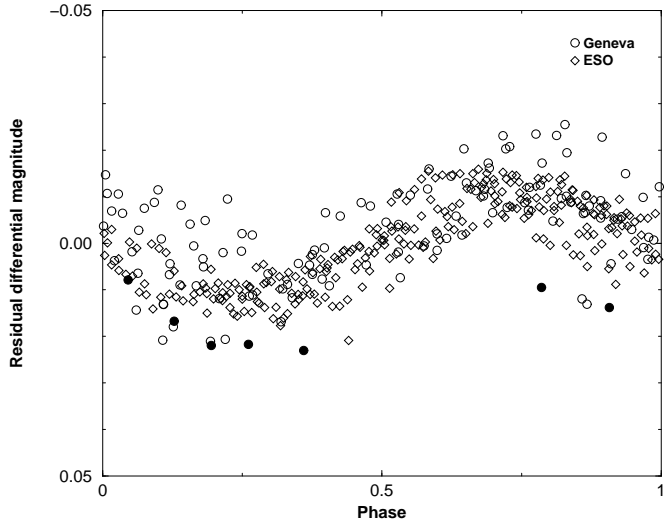}}
 \caption{Phase diagram for the HD~220392-data against the frequency of 5.52234 cpd 
 (after removal for the 4.67 cpd variation). Filled symbols represent
 the data on JD 2448518 }
\end{figure}

On Fig.~1, there is an additional peak at 9.37 cpd but a least squares fit gives 4.32 cpd
as a result for all datasets. However, evidence for this frequency is small as slight
changes in the datasets do not confirm its existence: e.g. if we remove the data of 
only one night of Geneva photometry (JD 8518) this peak disappears. 
The frequencies for a double-frequency fit were determined by minimization of a 
subset of 321 data with no quality degradation (i.e. we removed the data of JD 8518 and 
the high-airmass data obtained at ESO). The results of the final fit for all 396 data are 
found in Table~\ref{freq92}(c). The best match is obtained with the set of frequencies
(4.67439,5.52234). We present both mean light curves in Figs.~2 and~3: the first one 
shows all the data plotted against a frequency of 4.67439 cpd after having taken the 5.52 cpd 
variation into account while the latter one shows the same but this time against a 
frequency of 5.52234 cpd. The dispersion around both light curves is fair as it amounts to 
respectively 0.009 and 0.006 mag. Some 60\thinspace \% of the initial standard deviation
is thus removed.

\subsubsection{\protect\bigskip HIPPARCOS data}

The Hipparcos Epoch Photometry Catalogue contains 183 measurements of HD~220392 (HIP~115510). The
note in the Main Catalogue however mentions that the "data are inadequate
for confirmation of the period from Ref. 94.191" (ESA \cite{esa97}). The reason is that 
all the quality flags are equal to or larger than 16, meaning "possibly interfering
object in either field of view". The effec\-ti\-ve width of the aperture (called {\it 
Instantaneous-Field-of-View}) is 38 arcsec, so companions at angular separations between 
10 and 30 arcsec may interfere significantly during the measurement. We selected 177 data 
with a value of the quality flag not worse than 18, with a transit error on the (dc) 
magnitude not larger than 0.015 mag (2 data have not) and with good agreement between 
the (ac) and the (dc) magnitudes (1 datum has not) (ESA \cite{esa97}, Vol. 1, Appendix A). In addition, 
we had to eliminate one more datum, the brightest one. The mean of the remaining 
data is 6.204 mag with a standard deviation of 0.024 mag. Fourier analysis between 0. 
and 23. cpd shows a peak at 4.6743 $\pm$ 0.0001 cpd, i.e. the same main frequency as 
found in all former datasets. 
The corresponding phase diagram is illustrated in Fig.~4: the amplitude associated 
with $f_{1}$ is 0.013 mag large. The second frequency (5.52 or 6.52~cpd) is below detection: 
prewhitening for the main frequency still leaves a (large) dispersion of 
0.021 mag. A double-frequency simultaneous fit attributes an amplitude of 0.013 mag 
to $f_{1}$ but only 0.003 mag to $f_{2}$.

\begin{table}[tbp]
\setlength{\tabcolsep}{1.5mm}
\caption{Results of successive frequency analyses for HD 220392 (program PERIOD98).}
\label{freq92b}
\begin{center}
{\small
\begin{tabular}{|c|ccccc|}
\hline
Freq.  & Sept.'91 & June'90 & Oct.'92 & Sept.'91 & Total\struutup  \\
   &     & + Sept.'91 &  & + Oct.'92 &  \struutdown\\ \hline
Nights & 11  & 18  & 3  & 14  & 21  \struutup \\ 
Resol.  & 0.0523 & 0.0022 & 0.1082 & 0.0024 & 0.0012 \struutdown\\ \hline
\textit{{f$_{1}$}} & 4.674 & 4.679 & 4.661 & 4.664 & 4.674 \struutup \\ 
\textit{{(f$_{2}$)}} & (6.520) & (6.520) & (6.398) & (6.383) & 5.522 \\ 
\textit{{f$_{2}$}} & 5.523 & 5.520 & 5.536 & 5.523 & 5.522 \\
\textit{{f$_{3}?$}} & 4.320 & 4.316 & 4.320 & 4.321 & 4.320 \struutdown \\ \hline
\end{tabular} }
\end{center}
\end{table}

\begin{figure}[b]
 \label{pha510}
 \resizebox{\hsize}{!}{\includegraphics[height=5.5cm,angle=270]{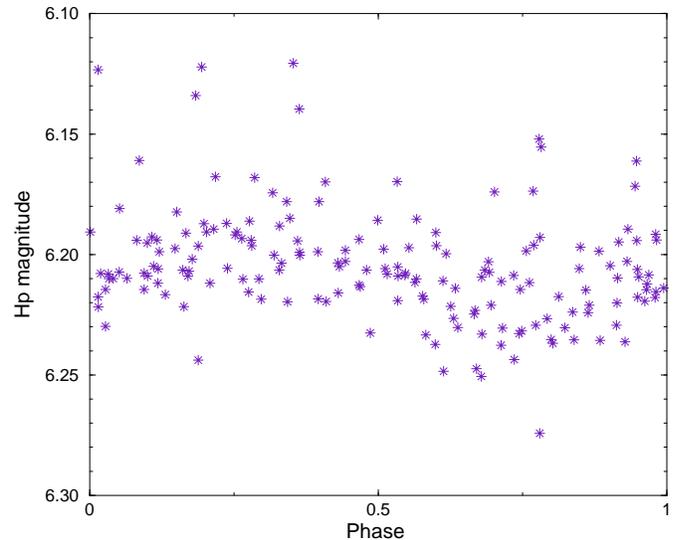}}
 \caption{Phase diagram for the HIP~115510-data against the frequency f$_{1}$ (4.67439 cpd) }
\end{figure}
\subsection{HD~220391}

\subsubsection{\protect\bigskip ESO and Geneva data}

245 observations were obtained during the last two seasons only, spanning
14\ nights. Again the data obtained on JD~8518 are conspicuously "low": the 
same effect as in the former data analysis was detected, implying an artificial 
increase in standard deviation of about 0.001 mag. We note the much smaller standard 
deviation of 0.0061 mag in the rest of the measurements. A frequency search was performed 
in a similar way as for HD~220392: only one peak at the frequency 0.42 cpd was found. 
However, the associated amplitude is below the expected noise level and the reduction 
of the standard deviation is very low (Table~\ref{freq91}). 
Additional observation campaigns should be undertaken to investigate the reality
of this frequency. 

\begin{table}[tbp]
\setlength{\tabcolsep}{1.0mm}
\caption{Results of the single-frequency fit for HD~220391 (program PERIOD98) }
\label{freq91}
\begin{center}
{\small
\begin{tabular}{|c|cccc|}
\hline
Campaign& Frequency & Semi-ampl. & Resid. $\sigma $ & Reduction \struutup\\ 
 & [cpd] & [mag] & [mag] & \% \struutdown\\ \hline
Sept. 1991 & 0.425 & 0.0034 & 0.0062 & 7.3 \struutup \\ 
Oct. 1992 & 0.422 & 0.0059 & 0.0041 & 33.0 \\ 
Total & 0.426 & 0.0050 & 0.0051 & 19.0 \struutdown \\ \hline
\end{tabular} }
\end{center}
\end{table}

\subsubsection{\protect\bigskip HIPPARCOS data}

The Hipparcos Epoch Photometry Catalogue contains 182 measurements of HD~220391 (HIP~115506). 
As in the first case, all quality flags are equal to or larger than 16. We selected 172 
data with a value of the quality flag not worse than 18, with a transit error on the (dc) 
magnitude not larger than 0.020 mag (6 data have not) and with good agreement between the 
(ac) and the (dc) magnitudes (3 data have not). The mean of these is 7.227 mag 
with a standard deviation equal to 0.026 mag. Fourier analysis between 0. and 23. cpd 
displays a peak at $\sim$ 11 cpd (with an associated amplitude of 0.013 mag!), 
an artefact frequency of order $2{\rm hr^{-1}}$, introduced by the rotation period of the 
satellite and very conspicuous in the spectral window Fouriergrams.

\section{Astrophysical considerations}\label{sec:nat}

\subsection{The nature of the association}\label{sec:ass}

\begin{table}[h]
\caption{Physical parameters for HD~220392/1}
\label{phys}
\begin{center}
{\small
\begin{tabular}{|c|c|c|c|}
\hline
Identifiers & HD & 220392 & 220391 \struutup \\ 
 &  Hip & 115510 & 115506 \\ 
 & CCDM & 23239-5349A & 23239-5349B \struutdown \\ \hline
Sp. Type &  & F0IVn & A9Vn \struutup\\ 
$\mathrm{m}_{v}$ & mag & 6.124 $\pm $ 0.014 & 7.103 $\pm $ 0.007 \\ 
U & mag & 1.608 & 1.538 \\ 
V & mag & 0.647 & 0.662 \\ 
B$_{1}$ & mag & 0.958 & 0.954 \\ 
B$_{2}$ & mag & 1.422 & 1.426 \\ 
V$_{1}$ & mag & 1.364 & 1.379 \\ 
G & mag & 1.769 & 1.790 \\ 
d & mag & 1.314 & 1.259 \\ 
m2 & mag & -0.491 & -0.494 \\
B$_{2}-$V$_{1}$ & mag & 0.058 & 0.047 \\ 
$M_{\mathrm{V}^{(1)}}$ & mag & +0.83 $\pm $ 0.15 & +1.62 $\pm $ 0.15 \\ 
$M_{\mathrm{bol}}$ & mag & +0.87 $\pm $ 0.15 & +1.66 $\pm $ 0.15 \\ 
log$T_{\mathrm{eff}}$ & K & 3.856 $\pm $ 0.008 & 3.867 $\pm $ 0.009 \\ 
$\lbrack M/H]$ & dex & -0.05 $\pm $ 0.09 & -0.12 $\pm $ 0.10 \\ 
log g & dex & 3.77 $\pm $ 0.07 & 4.06 $\pm $ 0.07 \\ 
$\mathcal{M}$ & $\mathcal{{M}_{\odot }}$ & 2.3 $\pm $ 0.2 & 1.8 $\pm $ 0.2 \struutdown\\ \hline
$vsin{\it i}$ & km$s^{-1}$ & 165 & 140 \struutup\\ 
$\pi _{Hip}$ & mas & 6.79 $\pm $ 1.43 & 9.19 $\pm $ 2.44 \\ 
$\pi _{phot}$ & mas & 8.7 $\pm $ 1 & 8.0 $\pm $ 1 \\ 
$M_{\mathrm{V}^{(2)}}$ & mag & +0.28 $\pm $ 0.46 & +1.92 $\pm $ 0.58 \\ 
$\mu _{\alpha }^{\ast }$ & ''/yr & 0.073 & 0.071 \\ 
$\mu _{\delta }$ & ''/yr & -0.035 & -0.033 \struutdown\\ \hline
\end{tabular}
\\[0pt]
}
$^{(1)}${\small from the Geneva Photometry calibration} \\[0pt]
$^{(2)}${\small based on the Hipparcos parallax }\\[0pt]
\end{center}
\end{table}

From the mean colour indices in the Geneva Photometric System and the corresponding
calibrations for A-F type stars (Hauck \cite{hau73}, K\"{u}nzli et al. \cite{kun97}), 
we derive the physical parameters
presented in the upper part of Table~\ref{phys}. An estimation of the masses
is obtained utilising the calibrations from Kobi \& North~(\cite{kob90}) and
North~(private comm.). Spectral types were determined by Gray \& Garrison~(\cite{gra89}).
Rotational velocities are from Levato~(\cite{lev75}). Bolometric corrections have
been taken from Flower~(\cite{flo96}). In addition, we have the Hipparcos
trigonometric parallaxes and proper motions, useful to establish the nature
of the association between both stars: the values of the parallaxes differ
by only 1 to 1.5 $\sigma _{\pi }$ and the resemblance of the proper motions
is striking (bottom part of Table~\ref{phys})(ESA \cite{esa97}). A very small
relative proper motion of magnitude 0.0029\arcsec/yr in the direction of 315$%
^{\circ }$~accompanies the large angular separation of 26.5\arcsec, which is the
reason of its classification as a common proper motion pair. For this reason 
and because both parallaxes are in reasonable agreement, the physical 
association of the pair is probable (van de Kamp \cite{vdk82}). The fact that both 
stars share the same location in space augments the probability that they were formed 
at the same time from the same parent cloud. 
Adopting the mean of both values as the system's parallax ($\pi _{\mathrm{AB}}$ = $7.99\pm 
2.~{\rm mas}$, in good agreement with $\pi _{\mathrm{phot}}$), we obtain a real separation of
the order of 3300 AU between the two components (neglecting the $\Delta \pi$
effect). For a mass sum of 4.1 M$_{\odot }$, the orbital period is very long, 
$\approx $ $10^{5}$ years. Both stars also share the same {\it projected} rotational velocity. 
We checked for radial velocity data as a further evidence of the wide association (i.e. 
we expect a small radial velocity difference). Grenier et al. (\cite{gre99}) 
published radial velocities for both stars only very recently: they determined 17.2 $\pm$ 0.69 km/s for 
HD 220392 and 10.75 $\pm$ 4.06 km/s for HD 220391 (while Barbier-Brossat et al. \cite{bar94} listed 
+6.7 km/s for HD 220392). Thus, not only is there a good agreement between both values, in addition 
it seems that component B has a variable radial velocity (No further conclusion can be drawn 
for the latter component as this is based on three measurements only).
Again making use of the Hipparcos parallax and of the definition of distance 
modulus, one can derive an absolute magnitude, $M_{\mathrm{V}^{(2)}}$, but - due to 
the relative error of 20-25\% on the parallaxes - the absolute magnitudes thus 
derived are too imprecise.

We give preference to the absolute magnitudes derived from the photometric
calibration, $M_{\mathrm{V}^{(1)}}$, to fit a model of stellar evolution of
solar chemical composition (Schaller et al. \cite{sch92}) in a theoretical H-R
diagram. The same isochrone with an estimated age of $\approx $ $10^{9}$
years for the system appears to fit both stars well (Fig.~5), as was also
verified by Tsvetkov (\cite{tse93}). We conclude that both stars form a common 
origin pair and probably even a true binary system.

\begin{figure}[tbp]
{\normalsize \label{iso} }
 \resizebox{\hsize}{!}{\includegraphics[width=8.5cm,angle=270]{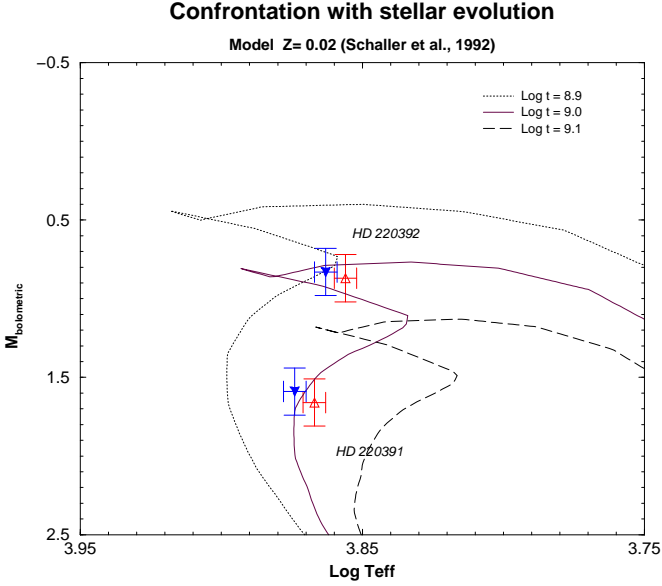}}
 \caption{Isochrone fit in the H-R diagram (Z=0.020). Filled symbols illustrate the locations
 after the correction for rotation }
\end{figure}

\subsection{The effects of rotation}\label{sec:rot}

In this section we want to investigate whether rotation could have an influence
on the derived physical quantities from Table~\ref{phys} and on the previously 
determined age and evolutionary phases. Both stars indeed seem to present rapid rotation 
and their photometric indices might be affected by the rotation effects such as
described by P\'erez Hern\'andez et al. (\cite{per99}) (hereafter PH99). In some cases these effects 
appear to be larger than the errors from the calibration: corrections for rotation 
have been considered by Michel et al. (\cite{mic99}) when analysing several
fast rotating $\delta$ Scuti stars of the Praesepe cluster. 

We recall here that the calibration of the multicolour Geneva colour indices in terms 
of various physical stellar parameters rests on a large sample of stars with well known 
spectroscopic characteristics (i.e. with known abundances, vsin{\it i}, spectral classification, 
etc) that have been measured in this photometric system. Such calibrations are therefore
in the first place empirical (Golay \cite{gol80}). They are based on real stars
and do not exactly correspond to non-physical objects (such as zero-rotating stars).
Spectroscopically calibrated parameters (such as T$_{\rm eff}$ and log g) will not
suffer too much from the effects of rotation however, mainly because slow rotators
will preferentially be chosen as reference objects because of a higher precision
of the stellar parameters. On the other hand, one must also recall that the mean 
rotational velocity of normal A9V and F0IV stars is $\simeq$ 130 km/s (Schmidt-Kaler \cite{sch82}). 
When we thus wish to correct the photometric indices for the effects of rotation, 
we will not need to apply the full range of proposed colour differences: the true correction
in the sense observed minus reference object will be smaller than the corrections computed by comparing a 
uniformly rotating model (represented by the observed star) to a non-rotating model (represented by the 
zero-rotation "copartner").

Because our targets have such similar properties, both in temperature and in projected 
rotational velocity, we determined the corrections for the secondary (a MS star)
and applied identical corrections to the more evolved component. To do this, we 
have estimated the break-up velocity and the rate of rotation for each of them. Using 
$\nu_{\rm break}= \frac{\Omega_{c}}{2\pi}$ and Eq.~(27) (PH99) with a polar radius 
R$_{p}$= 1.5 R$_{\odot}$ we find that $\nu_{\rm break,A} \simeq \nu_{\rm break,B}$ = 
41-42 $\mu$Hz. Since ${\rm vsin}{\it i}_{A}$ = 165 km/s
and ${\rm vsin}{\it i}_{B}$ = 140 km/s, we determine a rotation rate $\omega$ =
$\nu_{\rm rot}$/$\nu_{\rm break}$ smaller than 40\% for both. In addition, we may deduce
that the inclination is probably $> 30^{\circ }$. We applied the (excessive) colour
differences corresponding to $\omega$ = 50\%, i = $90^{\circ }$, log g$_{e}$ = 4.34 and 
log T$_{e}$ = 3.89, where \begin{displaymath} 
{\rm g}_{\rm e} \equiv \frac{\rm G \mathcal{M}}{{\rm R_{\rm p}}^2}
~{\rm and}~ {\rm T_{\rm e}}^4 \equiv \frac{\rm L}{4\pi\sigma{{\rm R_{\rm p}}^2}} 
\end{displaymath} (Eqs.~(21) and (22) in PH99).
The (over)corrected photometric parameters then are:\\
B$_{2}$-V$_{1}$=0.042,d=1.335, m$_{2}$=-0.489 for star A  and
B$_{2}$-V$_{1}$=0.031,d=1.281, m$_{2}$=-0.492 for star B.\\
The corresponding new locations of both stars in the H-R diagram are represented by 
the filled symbols in Fig.~5. The differences are of the order of the respective errors
but somewhat larger in T$_{\rm eff}$: 0.04-0.05 dex in log g (or -0.03 to -0.07 in M$_{\mathrm{bol}}$)
and 100~K in temperature. 
One may therefore safely state that the application of realistic corrections for the 
rotation of both stars does not really affect the previous conclusions 
re their physical properties, their age and evolutionary phases.

\subsection{The nature of the variability}\label{sec:var}

The mean (d,~B$_{2}$-V$_{1}$)-values place both stars well within the 
$\delta $ Scuti instability strip as observed in the Geneva Photometric
System. We note the interesting situation that two stars having such similar
characteristics behave quite differently from the variability point-of-view.
In the previous sections we have shown that the brighter component has a 
$\delta $ Scuti type of variability with a total amplitude of 0.05 mag while the
fainter component presents no short-period variability of amplitude larger
than 0.01 mag. What could the cause(s) be for this observed difference in variability? 
From the Geneva colour indices, it appears that the
brightest component has $\Delta d>0.100$, thus it is more evolved
than its companion. From the isochrone fit, one may also notice the
probable core hydrogen burning evolutionary phase of HD~220391 and the overall 
contraction or shell hydrogen burning phase of the
brighter component, HD~220392. Evolution appears here to be the most 
probable cause for the diversity in variability (in period and\//or amplitude) 
between the two stars.\\
Many $\delta$ Scuti stars are evolved objects (e.g. North et al. \cite{nor97}). 
It is further known that many $\delta$ Scuti stars in the advanced shell H burning 
stage showing single or double-mode pulsation with high amplitudes 
(semi-amplitude $\Delta V > 0.1$ mag) are confined to the cooler part of the 
instability strip (Andreasen \cite{and83}). In addition, these are slow rotators. 
We here have a case of an evolved $\delta$ Scuti star of {\it low} amplitude (with
a semi-amplitude of 0.014 mag if one considers {\it only} the main frequency - which is
disputable), presenting the signature of multiple frequencies and of {\it rapid} 
axial rotation. This is not surprising since low-amplitude pulsators cover the entire
instability strip (Liu et al. \cite{liu97}). We might conjecture that, in this case, 
the amplitude of the pulsation could be limited due to fast rotation. 
In fact, from the point-of-view of pulsation versus rotation, Solano \& Fernley
(\cite{sol97}) tend to believe that fast rotation favours the $\delta$ Scuti type
of pulsation. One could wonder why there is no evidence for short-period
variability of this type in the less evolved companion star. (A possible explanation
might be that the companion is an even faster rotator with a different 
(smaller) inclination than the more evolved star and that the amplitude(s) of the pulsation 
are further damped, possibly beyond photometric detectability.)   

Can we identify any pulsation mode for HD 220392? 
Expected values for a $\simeq$ 2 $M_{\odot}$ standard Population I model are 0.033 days (F), 
0.025 days (1H), 0.020 days (2H) or 0.017 days (3H) in the case of radial modes (l=0). 
For non-radial pressure modes (l=1), these values may be slightly larger: 0.036 days 
(f), 0.029 days (p1), 0.022 days (p2) \dots (Fitch \cite{fit81}; Andreasen et al. \cite{ane83}).
The physical parameters of 
Table~\ref{phys} may be used for the computation of the pulsation constant Q:\newline
\hspace*{0.5cm} log Q = log($f^{-1}$) + 0.5 log($M/M_{\odot }$) + 0.3 $M_{%
\mathrm{bol}}$ + 3 log($T_{\mathrm{eff}}$) -12.697, where $f$ is the frequency in cpd.\newline  
The propagation of errors shows that the error on the pulsation constant is
of order 0.003 days (0.07 on $\Delta (log Q)$). The results are given in
Table~6. The values thus computed are on the high side for a definitive mode
identification: one could draw the conclusion that the frequency 
$f_{2}$ possibly corresponds to the fundamental radial mode (F). We wish to
remark that non-radial g modes as well as undetected binarity are possible
reasons for higher values of Q (There is however no indication for the latter
from the Hipparcos results). The frequency ratio f$_2$/f$_1$, 0.84,
is not very helpful in this case.
We stress the fact that additional photometric observations for this interesting 
couple of stars are highly recommended. The obtained data are not sufficiently 
numerous to allow unambiguous solutions nor to solve for the multiple frequencies. 
Radial velocities would be needed too. 

\begin{table}[h]
\setlength{\tabcolsep}{1.5mm}
\caption{ Q-values for HD~220392}
\label{qval}
\begin{center}
{\normalsize \vspace{2mm} \small
\begin{tabular}{|c|cccc|}
\hline
\multicolumn{1}{|c|}{Identifier} & \multicolumn{1}{c}{Frequency} & 
\multicolumn{1}{c}{log Q} & \multicolumn{1}{c}{Q} & \multicolumn{1}{c|}{
Comment \struutup} \\ 
\multicolumn{1}{|c|}{} & \multicolumn{1}{c}{[cpd]} & \multicolumn{1}{c}{
[days]} & \multicolumn{1}{c}{[days]} & \multicolumn{1}{c|}{%
\struutdown} \\ \hline
HD 220392 & 4.674 & -1.358 & 0.044 $\pm$ 0.003 & \struutup \\ 
& 5.522 & -1.431 & 0.037 $\pm$ 0.003 & F? \\
& (6.522) & -1.501 & 0.032 $\pm$ 0.003 & (F?) \struutdown \\ \hline
\end{tabular}
}
\end{center}
\end{table}

\section{Conclusion}\label{sec:con}

Binary and multiple systems with pulsating variable components offer a unique
opportunity of coupling the information obtained by astrometric means (association
type - parallax - total mass) to the astrophysical quantities gained from the photometry
/spectroscopy (luminosity ratio - colours - pulsation characteristics)(see Lampens \& 
Boffin \cite{lam00} for a review of $\delta$ Scuti stars in stellar systems).
The detailed investigation of the differences in variability and simul\-ta\-neous\-ly in physical 
properties between two components of a binary system may provide clues with respect to 
the pulsation: differences in origin and age can be ruled out as well as differences in 
overall chemical composition. Stronger constraints exist for the determination of the 
position of the components in the H-R diagram, there is therefore less ambiguity in 
determining the evolutionary status and the mass than in the case of single variable stars.
This is important when one of the components is located in the zone where evolutionary 
tracks are bent (e.g. near the end of the core hydrogen burning phase). 

A relevant question is what factors determine the pulsation characteristics (the 
amplitudes and the modes) in the $\delta$ Scuti instability strip? 
We addressed this from the point-of-view of two bright A/F-type stars
that are both located in the $\delta$ Scuti instability strip and that are
shown to be physically associated, i.e. they either form a common origin pair
or they are the components of a true wide binary system. In this case,
evolution (and mass) is the most pronounced physical difference between 
both stars and it is very probable that this is the cause for the observed difference
in variability behaviour. Further observations are needed, the more that, since there 
is no evidence for any metal lines in the spectra, a comprehensive variability analysis
of this system might also help explaining the presence of non-variable, non-metallic 
stars in the instability strip. 

In the light of the discussion by Solano \& Fernley (\cite{sol97}) on the relation between 
rotational velocity and amplitude, we noted the remarkable similarity of the
projected rotational velocities: both stars are rather fast rotators. If 
fast rotation favours pulsation of the $\delta$ Scuti type, we expect 
to find short-period variability for the B-component as well! Since it is 
less evolved than its brighter companion, smaller amplitudes are expected.
This is another reason why intensive monitoring of this southern system is certainly
worthwhile. In our example it was very easy to identify the short-period
pulsating component and the information obtained from the astrometry could be
coupled to the astrophysical parameters of each component individually. Even
better would be to investigate these characteristics in a close visual
binary for which information on the orbital motion can also be derived. This
will allow to obtain a direct estimation of the stellar mass, independent
from the choice of modelisation. The derivation of the pulsation constant
will be more straightforward (the error on the mass defines the accuracy of
Q). More cases like this one should be looked into (see Frandsen et al. \cite{fra95}).

With this application in mind, we made a crossidentification between
the Annex of Variable Stars and the Annex of Double and Multiple Stars from
the Hipparcos Catalogue (ESA \cite{esa97}). Some 2500 systems with at least one
variable component have been identified. But the description of the
variability or the light curve in the Annex always refer to the combined
magnitudes. Additional observations should help identify which component is
variable and which are the binaries that offer the opportunity of
coupling the information obtained by astrometric means to the physical properties
in order to obtain a consistent picture of the system and its components.\newline

{\bf Acknowledgements}
We thank the Geneva team (especially Dr. G. Burki) for the telescope time put at our 
disposal in June~1990 and September~1991. Dr. Sperl is kindly acknowledged 
for making the programme Period98 available for this application. We appreciate
the help of Dr. L. Eyer (Geneva Observatory) in the selection of the Hipparcos Epoch 
Photometry data. We thank our colleague, Dr. J. Cuypers, for a critical reading and 
the referee, Dr. P. North, for constructive comments on this manuscript. The Geneva
data can be requested from the Geneva photometry team. The ESO data are available on 
request from the authors. This work has made use of the Simbad database, 
operated by the {\it Centre de Donn\'ees astronomiques de Strasbourg} (France). \newline


\begin{thebibliography}{}

\bibitem[1983]{and83} Andreasen~G.K., 1983, A\&A 121, 250
\bibitem[1983]{ane83} Andreasen~G.K., Hejlesen~P.M., Petersen~J.O., 1983, A\&A 121, 241
\bibitem[1994]{bar94} Barbier-Brossat~M., Petit~M., Figon~P., 1994, A\&AS 108, 603
\bibitem[1976]{blo76} Bloomfield~P., 1976, Fourier Analysis of Time Series: An Introduction, Wiley \& Sons (eds.), New York
\bibitem[1997]{esa97} ESA, 1997, The Hipparcos and Tycho Catalogues, ESA SP--1200
\bibitem[1981]{fit81} Fitch~W.S., 1981, ApJ 249, 218
\bibitem[1996]{flo96} Flower~Ph., 1996, ApJ 469, 355
\bibitem[1995]{fra95} Frandsen~S., Jones~A., Kjeldsen~H., et al., 1995, A\&A 301, 123
\bibitem[1980]{gol80} Golay~M., 1980, Vistas Astron. 24, 141
\bibitem[1989]{gra89} Gray~R.O., Garrison~R.F., 1989, ApJS 69, 301
\bibitem[1999]{gre99} Grenier~S., Burnage~R., Faraggiana, R., et al., 1999, A\&AS 135, 503
\bibitem[1973]{hau73} Hauck~B., 1973, Three-Dimensional Representation of A0-G5 Stars. In: Hauck~B., Westerlund~B.E. (eds.),
 Proc. IAU Symp. 54, Problems of Calibration of Absolute Magnitudes and Temperature of Stars. Reidel, Dordrecht, p.117
\bibitem[1990]{kob90} Kobi~D., North~P., 1990, A\&AS 85, 999
\bibitem[1997]{kun97} K\"{u}nzli~M., North~P., Kurucz~R.L., Nicolet~B., 1997, A\&AS 122, 51
\bibitem[1992]{lam92} Lampens~P., 1992, Delta Scuti Newslet. 5, 9
\bibitem[2000]{lam00} Lampens~P., Boffin~H.M.J., 2000, $\delta$ Scuti Stars in Stellar Systems: a Review. In: Breger~M., 
  Montgomery~M. (eds.), 6th Vienna Workshop in Astrophysics on Delta Scuti and Related Stars. Vienna (in press)
\bibitem[1975]{lev75} Levato~A., 1975, A\&AS 19, 91
\bibitem[1997]{liu97} Liu~Y.Y., Baglin~A., Auvergne~M., Goupil~M.J., Michel~E., 1997, ESA SP-402, 363
\bibitem[1999]{mic99} Michel E., Hern\'andez~M.M., Houdek G. et al., 1999, A\&A 342, 153
\bibitem[1997]{nor97} North~P., Jaschek~C., Egret~D., 1997, ESA SP-402, 367
\bibitem[1999]{per99} P\'erez Hern\'andez~F., Claret~A., Hern\'andez~M.M., Michel~E. 1999, A\&A 346, 586 (PH99)
\bibitem[1988]{ruf88} Rufener~F., 1988, Catalogue of stars measured in the Geneva Observatory
 Photometric System, 4th Ed., Geneva Observatory (ed.), Geneva 
\bibitem[1992]{sch92} Schaller~G., Schaerer~D., Meynet~G., Maeder~A., 1992, A\&AS 96, 269
\bibitem[1982]{sch82} Schmidt-Kaler~T.H., 1982, in: Landolt-Bornstein New Series 2b, Schaifers~K., Voigt~H.H. (eds.), Springer-Verlag
\bibitem[1997]{sol97} Solano~E., Fernley~J., 1997, A\&AS 122, 131
\bibitem[1998]{spe98} Sperl~M., 1998, Manual for Period98 : V1.0.4, {\it (http://dsn.astro.univie.ac.at/period98)}
\bibitem[1993]{tse93} Tsvetkov~Ts.G., 1993, Ap\&SS 208, 285
\bibitem[1982]{vdk82} van de Kamp,~P. 1982, Evolutionary Trends in Wide Binaries. In: Kopal~Z., Rahe~J. (eds.), Proc. IAU Coll. 69, 
 Binary and Multiple Stars as Tracers of Stellar Evolution. Reidel, Dordrecht, p.81
\bibitem[1997]{wor97} Worley~C., Douglass~G., 1997, A\&AS 125, 523

\end{thebibliography}
\end{document}